# Effect of Cold Sintering Process (CSP) on the Electro-Chemo-Mechanical Properties of Gd-doped Ceria (GDC)

Ahsanul Kabir[1,3], Daoyao Ke[2], Salvatore Grasso[2], Benoit Merle[3] and Vincenzo Esposito[1*]

[1]Department of Energy Conversion and Storage, Technical University of Denmark, Frederiksborgvej 399, Roskilde 4000, Denmark

[2]Key Laboratory of Advanced Technologies of Materials, Ministry of Education, School of Materials Science and Engineering, Southwest Jiaotong University, Chengdu, 610031, China

[3]Materials Science & Engineering I and Interdisciplinary Center for Nanostructured Films (IZNF), University Erlangen-Nürnberg (FAU), Cauerstr. 3, 91058 Erlangen, Germany

*Corresponding Authors: E-mail: ahsk@dtu.dk, vies@dtu.dk

## Abstract

In this report, the effect of the cold sintering process (CSP) on the electro-chemo-mechanical properties of 10 mol% Gd-doped ceria (GDC) is investigated. High purity nanoscale GDC powder is sintered via a cold sintering process (CSP) in pure water followed by post-annealing at 1000 °C. The resultant CSP ceramics exhibits high relative density (~92%) with an ultrafine grain size of ~100 nm. This sample illustrates comparable electrochemical properties at intermediate/high temperatures and electromechanical properties at room temperature to the sample prepared via conventional firing, *i.e.* sintering in the air at 1450 °C. Moreover, a large creep constant as well as a low elastic modulus and hardness are also observed in the CSP sample.

## 1. Introduction

Ceria is a functional metal oxide with a centrosymmetric cubic fluorite structure and excellent mechanical, electrical, chemical properties including high ionic conductivity, catalytic redox properties, hardness and chemical resistance [1][2][3][4]. Due to its multifold properties, the cerium oxide-based compound has attracted growing interest for a wide range of industrial applications such as electrolytes for solid oxide fuel cells, automobile catalysts, oxygen gas sensors, *etc.* [5][6]. Moreover, a recent investigation illustrates that ceria exhibits a non-classical giant electrostriction effect, with a magnitude two orders higher than classical prediction [7][8]. Ceria is normally doped with rare-earth aliovalent dopants such as $Gd^{3+}/Sm^{3+}$, leading to the formation of charge compensating extrinsic oxygen vacancies ($V_O^{\cdot\cdot}$) in the lattice. Additionally, cerium undergoes reduction reaction ($Ce^{4+} \rightarrow Ce^{3}$) at low oxygen partial pressure at high temperatures, resulting in a quasi-free localized electron and oxygen vacancy [9]. In general, the presence of oxygen vacancies in the lattice and their local configuration controls the material´s properties, howbeit with a strong dominance from final density and microstructure evolution. Being a high melting point material (~2400 °C), ceria usually requires a high sintering temperature of ~1400-1500 °C for complete densification [1][10]. Such a process is energetically demanding and the thermally activated diffusion can trigger undesired chemical reactions within the material and surrounding environment, including toxic fumes (NOx, CO, VOCs, etc.) and other solids byproducts [11][12][13]. In view of that, a drastic reduction of sintering temperature would be a real interest for technological applications. Numerous sintering techniques have been developed over the years such as two-step sintering (TSS), field-assisted sintering (spark plasma sintering/flash sintering), microwave sintering, *etc.* [14][15]. Another way to enhance densification is to use nanoparticles or/and sintering aids (liquid phase sintering) [4][16][17]. All these methods not only lessen the sintering temperature (~10–25% of regular temperature) but also alters materials' functionality to some extent. Recently, a novel low temperature (<300 °C) non-equilibrium sintering process named cold sintering process (CSP) has been developed by Randal *et al.* [18][19]. Simply stated, the CSP is a pressure-assisted liquid phase sintering process. In this method, ceramic powders are mixed with an aqueous solution (commonly water) as a transient solvent, resulting in the partial dissolution of powder particle and accordingly reduction of interfacial area. With sufficient heat (>100 °C) and external pressure (50–500 MPa), the liquid phase redistributes into the particle–particle contacts. This enables rapid mass flow as the liquid evaporates, eventually creates a supersaturated solution and precipitation, leading to minimize the free surface energy of the particles and form a dense compact [20][21]. The densification takes place within 1–60 minutes reaching nearly ~80–98% of the theoretical density. So far, a broad variety of materials with different functions including ferroelectrics, semiconductors, ionic electrolytes, and thermoelectrics have been successfully prepared via CSP [22]. However, for many materials, *e.g.* 8YSZ a second thermal step (lower than conventional temperature) might be needed to accomplish complete densification and stabilize the material for high-temperature applications [23][24]. Intrinsic parameters such as composition, dopants, solvent type, solubility as well as external factors, *e.g.* pH, sintering pressure/time/temperature profile significantly influence the CSP mechanism and kinetics [25]. The design of the CSP method is simple, requiring a steel die, press, and heat controller [26]. Despite the large technological relevance, until now, little attention has been paid to the viability of applying cold sintering to the ceria-based compound.

The main aim of the current work is to examine the effect of applying the cold sintering process on 10 mol% Gd-doped ceria (GDC), a material that is used extensively for commercial electro-ceramics applications. For this purpose, the GDC sample is prepared via the CSP process, followed by an additional post-heat-treatment. The electrical, mechanical and electromechanical properties are highlighted, and the relationships interlinking microstructure and properties are derived. Moreover, the results are compared with the samples sintered either via a conventional treatment in the air (*i.e.* free sintering) [12] or a field-assisted spark plasma sintering (SPS), using the same starting materials.

## 2. Experimental Procedure

### 2.1 Sample Preparation

Commercial 10 mol% Gd-doped ceria (GDC10) powders (Rhodia, France) with specific surface area ≈ 35 m$^2$/g, as characterized by Braunauer-Emmett-Teller (BET) method (Quantachrome Autosorb-1-MP, Germany) were used in this experiment. The step of the cold sintering process is schematically shown in **Fig. 1**. To begin with, 1 g GDC powder and 0.4 g deionized water were mixed in a centrifuge tube using a vortex mixer. The resulting paste was then loaded in a 15 mm inner diameter steel die and pressed under a uniaxial pressure of 300 MPa. Afterward, the temperature was gradually raised from room temperature up to 200 °C and subsequently held for 30 minutes. The experimental density of the as-prepared green pellet was roughly 75%. With the aim of improving further densification, the sample was post-annealed at 1000 °C for 1 h with a heating rate of 10 °C/min. For the property comparison, one sample was conventionally dry-pressed under a uniaxial pressure of 150 MPa for 30 s and freely sintered at 1450 °C for 1 h in air. Another sample was consolidated by the SPS (Dr. Sinter Lab 515S, Japan) under a high vacuum (≤ 6 · 10$^{-6}$ Torr) at 1100 °C with a constant uniaxial pressure of 50 MPa for 5 min dwelling. To eliminate the chemical reduction, the SPS sample was re-oxidized at 800 °C for 10 h.

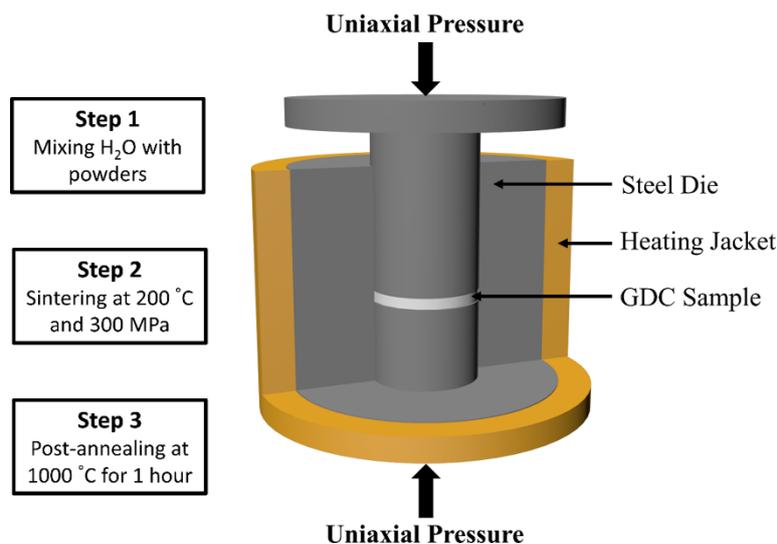

**Figure 1:** Schematic illustration of the steps in the cold sintering process (CSP) for GDC ceramics.

## 2.2 Materials Characterization

The experimental density of the samples was measured by the Archimedes method in distilled water. The crystallographic phase purity and microstructure were investigated by the X-ray diffraction (XRD) method (Bruker D8, Germany) and a high-resolution scanning electron microscope (SEM) (Zeiss Merlin, Germany), respectively. The electrochemical properties were analyzed by electrochemical impedance spectroscopy (EIS) Solarton 1260 (UK) in a temperature range of 250–500 °C for a frequency distribution of 0.01 Hz to 10 MHz with an alternative voltage of 100 mV in air. Prior to measurement, the samples were coated with silver paste in a symmetric configuration and dried at 600 °C for 15 min. The electromechanical property is characterized in a nanovibration analyzer with a single-beam laser interferometer (SIOS, Germany). The alternate electric fields were applied by varying frequencies and subsequently, vertical displacements/strains were measured. The mechanical properties of the samples were investigated by the nanoindentation technique (KLA G200, USA) with a Berkovich indenter at room temperature. The creep measurements were based on a trapezoidal load-hold-unload scheme [27]. The sample was loaded at a rate of 15 mN/s (fast loading mode) until a value of 150 mN – alternatively a depth of 1000 nm – was reached. The load was subsequently held constant for 20 s (see **Fig. 5a**). During the hold segment, the creep relaxation of the material was recorded as a progressive increase in displacement. Unloading was performed at the same rate as loading and served for determining the elastic modulus and hardness of the material using the Oliver-Pharr method [28]. For evaluating the local strain rate sensitivity, the indentation strain rate was varied abruptly between 0.001 $s^{-1}$ and 0.1 $s^{-1}$ during the measurement [29], and the strain rate sensitivity was estimated from the resulting jump in hardness [27][30]. The experimental values reported for each sample are based on measurements at >8 different locations.

## 3. Results and Discussion

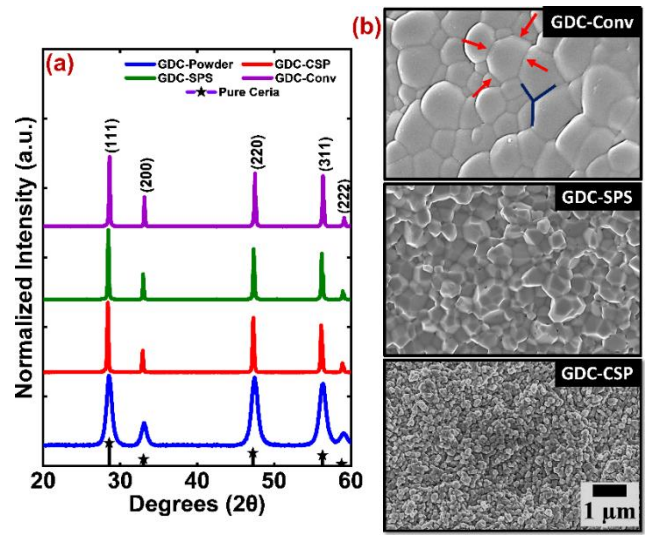

**Figure 2:** (a) X-ray diffraction (XRD) pattern of 10 mol% gadolinium doped ceria (GDC) powder and sintered pellets. (b) Scanning electron microscope (SEM) images of the GDC pellets, respectively sintered in conventional firing at 1450 °C for 1 h, SPS at 1100 °C for 5 min, and CSP at 200 °C 300 MPa, post-annealing at 1000 °C for 1 h.

**Fig. 2** reveals the crystallographic and microstructural properties of the samples. The XRD reflection peaks of the pattern are indexed with the theoretical pattern of pure ceria (ICSD code 251473), showing a single-phase cubic fluorite structure (see **Fig. 2a**). Bearing in mind the resolution of the XRD method, no other secondary phase was evident. The outcome is in accordance with the previous literature [1][31]. For the sintered pellet, the peaks became narrow and sharp due to particle size coarsening. The SEM micrographs of the sintered pellet are presented in **Fig. 2.b**. The microstructure is highly compact and is in agreement with values of experimental density (> 97% of theoretical density) for SPS and Conv sample. The GDC-CSP sample is less dense and consists of ca. 8-10 % of residual porosity, as measured with the ImageJ tool [32]. As expected, the GDC-Conv sample displays on average a large micron-size grain, with an inhomogeneous distribution. Most of the big equiaxed cubic grains (fully relaxed) have the thermodynamical equilibrium shape at the tripe point (blue lines in **Fig. 2b**) with a small residual grain boundary curvature (red arrows in **Fig. 2b**). On the other hand, a non-relaxed homogenous microstructure is recorded for both GDC-SPS and GDC-CSP samples. These samples demonstrate typical polygonal-shaped nano-grains with an average size of about 300 ± 30 and 100 ± 20 nm, respectively.

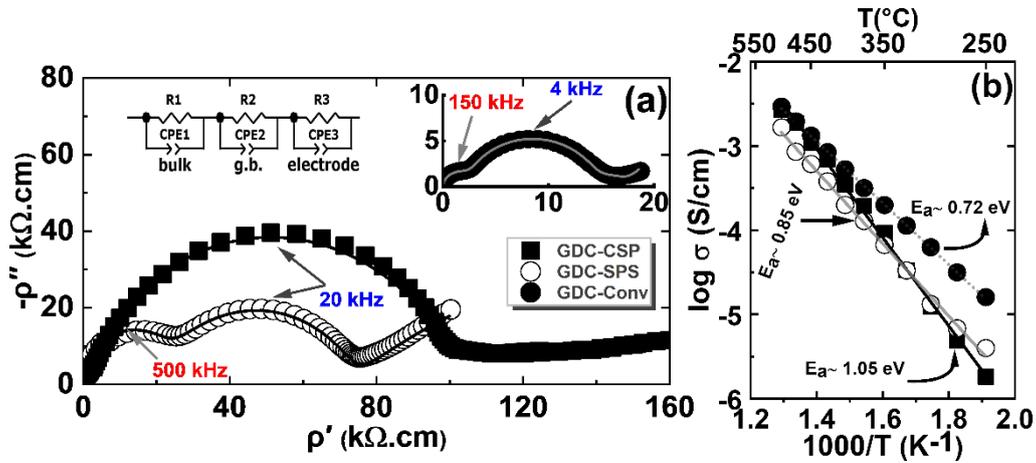

**Figure 3:** Typical geometry normalized (a) Nyquist plots (ρ' vs ρ") characterized by frequencies, recorded at 300 °C in air for the GDC pellets. (b) Arrhenius plot for the estimation of total electrical conductivities of the GDC samples.

**Fig. 3a** illustrates the geometry normalized Nyquist plots (ρ' vs ρ") measured at 300 °C in ambient air. An equivalent circuit model of the *RQ* element was used to fit these plots. *R* and *Q* are denoted respectively as resistor and constant phase element, where capacitance $C = (R^{1-n}Q)^{1/n}$. The GDC-Conv and GDC-SPS samples exhibit two well-defined semi arcs that refer to the high and intermediate frequency associated bulk and grain boundary impedance, respectively [33]. In contrast, a single semicircle is observed in the GDC-CSP sample, ascribing to superimposed bulk and grain boundary contribution. This kind of response was previously acknowledged for the nanocrystalline ceria compound, which has a similar space charge width and grain size [34]. The low-frequency tailed feature of the impedance spectra is referred to as an electrode/material interface polarization mechanism, which is not relevant in the analysis of the material´s electrochemical properties. According to bricklayer model, conduction property is principally dominated by the blocking effects at the grain boundary, which is characterized by the grain boundary blocking factor ($\alpha_{gb}$) where $\alpha_{gb} = \frac{R_{gb}}{R_{bulk}+R_{gb}}$ [35]. The estimated grain boundary-blocking factor ($\alpha_{gb}$) is smaller in GDC-SPS ($\alpha_{gb} = 0.60$) sample than of GDC-Conv ($\alpha_{gb} = 0.85$). Due to the evolution of a single semicircle in GDC-CSP material, it was not quantitatively possible to estimate the blocking factor. Taking into account the relaxation frequency of this sample, it can be concluded that semicircle in GDC-CSP is mastered resolutely by the grain boundary effect. Furthermore, it can be seen that the total resistivity of GDC-CSP is at least 5 times higher than the GDC-Conv sample. The reason for this behavior is due to the development of large blocking effects in the CSP composition that is strongly controlled by sintering characteristics *e.g.* sintering rate, final densification/microstructure, residual pore, *etc*. **Fig. 3b** shows temperature (T) dependence of the total electrical conductivities (σ) of the GDC samples in the Arrhenius equation:

$$\sigma T = \sigma_0 \exp\left(\frac{-Q}{kT}\right) \quad (1)$$

Where Q is the activation energy, $\sigma_0$ and k denotes as pre-exponent factor and Boltzman constant, respectively. As observed, GDC-SPS and GDC-CSP samples have comparable conductivities at low temperatures (<350 °C), however one order of magnitude lower than the GDC-Conv sample. Whereas at higher temperatures (>400 °C), the conductivity value of the GDC-CSP surpasses GDC-SPS, with values equivalent to the GDC-Conv sample. The activation energy is highest for the GDC-CSP sample which is associated with the large blocking factor effect. This sample also has a lower relative density than counterparts.

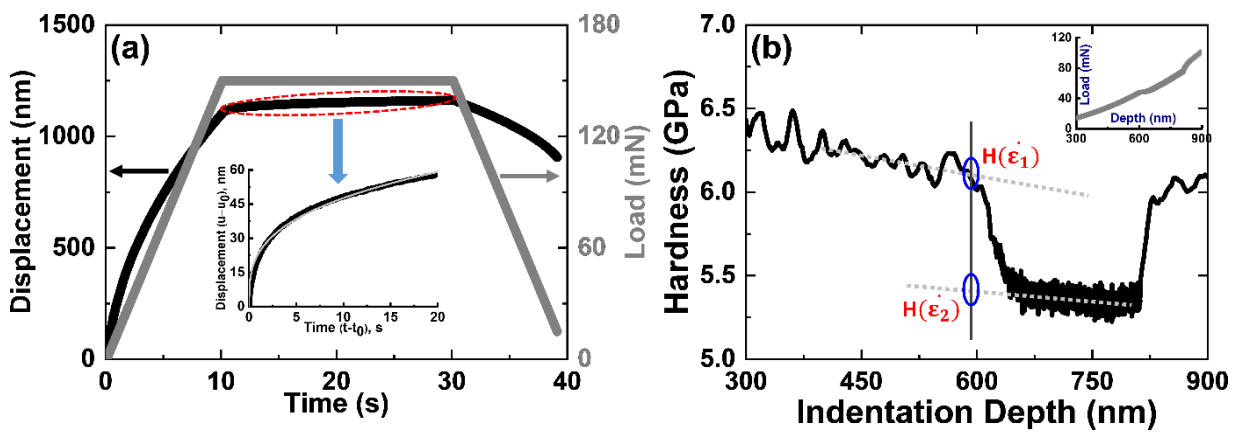

**Figure 4:** (a) A distinctive load/displacement–time curve in the nanoindentation measurement for the GDC-CSP sample (fast loading mode). The inset shows the displacement as a function of holding time under constant load, representing a primary creep behavior. (b) Indentation-depth dependence hardness during the strain rate sensitivity measurement.

The mechanical characterization of the samples is performed by the nanoindentation at room temperature. The applied indentation depth of ca. 1000 nm corresponds to an equivalent contact diameter of ca. 5500 nm, which means that multiple grains are tested at a time. Thus, the experimental results are to be considered as macroscopic mechanical properties [36]. In all measurements, the load–displacement curves were smooth and no pop-in events were observed (data not shown here). A typical load/displacement–time curve is shown in **Fig. 4**. All tested samples undergo a gradual increase in displacement – *i.e.* evidence a creep behavior – during the hold step of the measurement (see **Fig. 4a**). In cerium oxides, creep is predominantly driven by the rearrangement of elastic dipoles (lattice point defects) due to anisotropic stress [37][38]. The creep-driven increase in displacement with dwelling time can be described by the following empirical formula:

$$(u - u_0) = A\,(t - t_0)^m \qquad (2)$$

Where $u_0$ is the initial displacement at the commencement of hold stage $t_0$, $A$ is the creep constant and *m* the fitting exponent. The values of these parameters are listed in **Table 1**. The magnitude of *A* remains fairly constant, irrespective of whether the creep segment is initiated upon reaching a given load or a given depth (compare "150 mN load fast" and "1000 nm fast"). Although not material parameters, the fitting parameters *m* and *A* are useful to perform qualitative comparisons between different materials. Significantly, the *A* value is higher in the GDC-CSP sample than others are. The larger creep is likely connected to the GDC-CSP sample due to ultra-fine grains with a large grain boundary area and residual porosity. In general, pore (void) assists lattice rearrangement by increasing the grain-to-grain contact areas, leading to decreased residual stress. On the other hand, a high density of grain boundaries potentially increases the magnitude of deformation by the grain boundary sliding mechanism [39][40]. Therefore, the creep analysis suggests that the grain boundary is not continuous in the CSP samples. The creep fitting parameter *m* varies between 0.3–0.4 and inversely correlates with the experimental *A* values. Moreover, every sample shows a comparable *A* value in different experiments, suggesting that the measurements are unaffected by the indentation size effect (ISE), as well as surface defects such as microcracks, roughness, porosity, *etc*. The measured elastic modulus (*E*) and hardness (*H*) values differ significantly between the samples (see **Table 1**). In basic terms, these values display a decreasing trend with decreasing grain sizes. For instance, the *E* and *H* values of the GDC-CSP sample shows a significantly lesser value (35-40% smaller) than of GDC-Conv. To analyze the size effect of the deformation behavior, the local strain rate sensitivity was estimated from two different indentation depths (**Fig. 5b**). The strain rate sensitivity (*s*) is determined from the following equation:

$$s = \frac{\ln H_2 - \ln H_1}{\dot{\varepsilon}_2 - \dot{\varepsilon}_1} \tag{3}$$

Where H is the hardness and ε the strain rate. The calculated strain rate sensitivity is constant and ~0.025 for all the samples, and therefore independent of either grain sizes or porosity.

**Table 1:** The measured creep property, elastic constant, and hardness of GDC pellets.

| Parameter | Sample ID | Creep Constant, A (nm · s$^{-m}$) | Creep Fitting Parameter, m | Elastic Modulus, E (GPa) | Hardness, H (GPa) |
|---|---|---|---|---|---|
| **1000 nm (Fast)** | GDC-Conv | 14.3 ± 1.8 | 0.40 ± 0.015 | 230 ± 11.5 | 8.4 ± 0.8 |
| | GDC-SPS | 15.9 ± 0.6 | 0.35 ± 0.015 | 195 ± 8.9 | 7.2 ± 0.5 |
| | GDC-CSP | 19.0 ± 1.3 | 0.30 ± 0.01 | 135 ± 1.8 | 5.4 ± 0.1 |
| **150 mN (Fast)** | GDC-Conv | 16.9 ± 2.5 | 0.34 ± 0.012 | 225 ± 12.3 | 8.3 ± 0.7 |
| | GDC-SPS | 18.4 ± 1.1 | 0.34 ± 0.012 | 195 ± 4.9 | 7.3 ± 0.2 |
| | GDC-CSP | 21.9 ± 0.6 | 0.30 ± 0.004 | 135 ± 2.4 | 5.3 ± 0.1 |

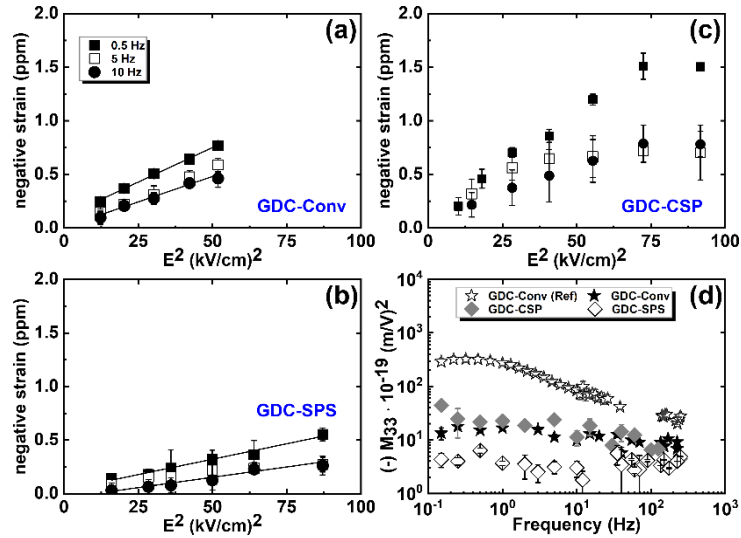

**Figure 5:** (a), (b) and (c) The electrostrictive response of GDC samples with a response to the external electric field at frequencies 0.5–10 Hz, electrode material: Au. (d) The electrostriction strain coefficient ($M_{33}$) as a function of frequencies, in between 0< f< 250 Hz. The results are compared with a reference GDC-10 sample (non-sputtered electrode), sintered at 1450 °C for 10h [31].

The electrostrictive properties of the GDC pellets are presented in **Fig. 5**. All these samples responded at the second harmonic of the applied electric field which shows a negative strain parallel to the field direction, confirming their electrostrictive behavior. As noted, the GDC-SPS display minimum strain whereas, comparable performance is observed for GDC-Conv and GDC-CSP samples, at certain fields and frequencies. For example, the strain value of the CSP and Conv sample is ~0.60 and ~0.45 ppm at 6 kV/cm and 5 Hz, respectively. Moreover, the magnitude of strain decreases gradually with increasing frequencies and strain tends to saturate at higher electric fields. Analogous behavior is detected in previously published reports [31][41]. The electrostriction strain coefficient ($M_{33}$) slowly decreases (non-Debye type) with increasing frequencies and are displayed in **Fig. 5d**. It has been reported that a high blocking factor strongly influences (increases) the electrostriction coefficient, especially at the low-frequency regime [31]. In view of this fact, the lower coefficient at the GDC-SPS sample is associated with the low-blocking effect whereas higher value in counterparts (CSP and Conv) is due to the presence of a large blocking factor. Moreover, the measured value is still one order lower magnitude in the low-frequency regime than of Ref. sample measured in the non-sputtered electrode, further emphasizing that the electrode is also a critical factor to measure this property.

## 4. Conclusion

In summary, this work reveals for the first time the effect of the cold sintering process (CSP) on the properties of 10 mol% Gd-doped ceria (GDC) ceramics, with a direct property comparison to the sample produced via the SPS and conventional thermal protocol. The CSP sample displays non-equilibrium nanoscale grains with a nearly homogeneous distribution, whereas relaxed grains are observed for the conventional material. The GDC-CSP develops a large blocking effect at the grain boundaries, which decreases electrical conductivity at low temperatures. Moreover, conductivity increases with temperature and matches the conventional sample at intermediate temperature. The elastic modulus and hardness are small in GDC-CSP samples due to residual porosity, which slightly increases the creep behavior. Surprisingly, this sample shows an equivalent electromechanical performance, in comparison with GDC-Conv. In summary, we conclude that CSP matches the performance of the conventional sample and can be applied to a wide range of applications.

## Acknowledgments

This research was supported by DFF-Research project grants from the Danish Council for Independent Research, Technology and Production Sciences June 2016 (GIANT-E, 48293), European H2020-FETOPEN-2016-2017 project BioWings grant number 801267 (partially), and the Center for Nanoanalysis and Electron Microscopy (CENEM) and the Interdisciplinary Center for Nanostructured Films (IZNF) at University Erlangen-Nürnberg (FAU). S.Grasso. was supported by the Thousand Talents Program of China and Sichuan Province.